# A Fundamental, Relativistic, and Irreversible Law of Motion: A Unification of Newton's Second Law of Motion and the Second Law of Thermodynamics

Randy Wayne

Department of Plant Biology, Cornell University, Ithaca, New York 14853 USA

**Abstract:** Temperature is an outsider in the laws of motion given by Newton and Einstein and this oversight is the source of the predictions of time-reversal-invariance made by these two great systems of motion. By taking into consideration Planck's law of blackbody radiation and the Doppler effect, in thinking about Maxwell's electromagnetic wave equation, I have shown that photons, in the environment through which any charged particle moves, act as a source of temperature-dependent friction on everything from elementary particles to galaxies. Because the optomechanical friction is universal and inevitable, no real systems are conservative, and temperature can no longer be an outsider in a fundamental and irreducible law of motion. I have defined the change of entropy ($\varDelta S$) in irreversible systems at constant temperature in terms of the optomechanical friction. The Second Law of Thermodynamics, which states that $\varDelta S > 0$ for spontaneous processes, is explained by electromagnetic interactions between charged particles and the Doppler-shifted photons through which they move as opposed to chance and statistics. $\varDelta S$, as defined here, is not subject to Poincaré's recurrence theorem. Consequently, the Second Law of Thermodynamics is shown to be a fundamental law rather than a statistical law. This result, which supports the idea that every instant of time is unique, is consistent with intuition and the routine experience of botanists.

*"Sometimes I dread the results of universal education. We become so indoctrinated with traditional ideas which may not be true, and we become so humble before the imposing prestige of accepted professional opinion, that we lose the power of independent creative thinking. We need a 'Society for the Preservation of Ignorance,' so that occasionally someone…may look at the world anew and tell us what he sees*[1]*."*

*--Arthur E. Morgan, 1937*

## 1. Introduction

Erwin Schrödinger[2] began his essay on "irreversibility" with the following words, *"It may seem an audacity if one undertakes to proffer new arguments in respect of a question about which there has been for more than eighty years so much passionate controversy, some of the most eminent physicists and mathematicians siding differently or favouring opposite solutions— Boltzmann, Loschmidt, Zermelo, H. Poincaré, Ehrenfest, Einstein, J. von Neumann, Max Born, to name only those who come to me instantly."* It probably seems even more audacious, perhaps even bodacious, for a botanist to proffer a new argument on irreversibility and the arrow of time, especially, when, currently *"the standard party line among theoretical physicists*[3]*"* such as Paul Davies[4], Julian Barbour[5], Brian Greene[6] and Sean Carroll[7,8] states that the arrow of time is an illusion since *"an examination of the laws of physics reveals only the symmetry of time*[9]*."* Brian Greene[6] emphasizes that *"the laws of physics that have been articulated from Newton through Maxwell and Einstein, and up until today, show a complete symmetry between past and future. Nowhere in any of these laws do we find a stipulation that they apply one way in time but not in the other. Nowhere is there any distinction between how the laws look or behave when applied in either direction in time. The laws treat what we call past and future on a completely equal footing. Even though experience reveals over and over again that there is an arrow of how events unfold in time, this arrow seems not to be found in the fundamental laws of physics."*

Nevertheless, while working as a botanist, or more specifically a biophysical plant cell biologist, patiently studying cells under a microscope and simply paying attention to the seemingly unremarkable processes that are accessible to everybody[10-22], I have gained a deep insight into viscous forces, friction, irreversibility, and the nature of reality. Consequently, I think that I have something valuable to say concerning irreversibility and why the arrow of time must be included in the fundamental laws of nature heretofore proffered by the theoretical physicists[10]. Indeed, much of what we call "physics" today resulted from the discovery and analysis of biological phenomena. For example, the transverse wave theory of light emerged from Thomas Young's studies of vision[23]; the discovery of Brownian motion grew out of Robert Brown's investigation of pollination[24]; the First Law of Thermodynamics resulted from Robert Mayer's observations of the color of blood[25]; the laws of solutions[26,27] came out of Wilhelm Pfeffer's [28] work on leaf movement and Hugo de Vries'[29] work on plasmolysis; and Poiseuille's law of laminar flow and Fick's Law of diffusion came from Jean Poiseuille's[30] study of blood flow in veins and Adolf Fick's[31] investigation of salt transport in kidneys, respectively.

In this paper, I present a new fundamental law of motion based on Newton's Second Law and a relativistic extension of Maxwell's electromagnetic wave equation that takes the Doppler effect into account. The new law is consistent with the principle of relativity and the invariance of the speed of light—the two postulates of Einstein's Special Theory of Relativity. However, in contrast to the laws of physics articulated by Newton, Maxwell, and Einstein, the new fundamental law of motion, which unifies dynamics and thermodynamics, is asymmetrical with respect to time, and is consistent with our everyday, commonsense, intuitive picture of the world that suggests that the orders of events exhibited during natural processes are asymmetrical with respect to the passage of time[32,33].

The asymmetrical arrow of time is supported by the empirical research of botanists who studied the fundamentally irreversible processes that occur in plants over time scales of femtoseconds[34,35], minutes, hours, days, and years[36-53] and even millions of years[54-60]. In fact, in order to initiate many of the natural processes involved in growth and development, the plants themselves utilize their own ability to tell time[61-67], and as an aside, the annual rings of long-lived trees that *"show the swing of Time's pendulum and put down a mark"* have been used as *"chronographs, recording clocks"*[68] by archeologists to determine the dates of historical events[68,69,70].

The life of a plant is fundamentally irreversible. When seeds germinate, they first produce roots and then light-seeking, photosynthetic leafy shoots[71]. The water- and nutrient-mining roots then grow though the soil parallel to the gravitational force, and the shoots, which grow toward the light and up[72], antiparallel to the gravitational force, eventually become reproductive[73]. In angiosperms, which are known as the flowering plants, the reproductive shoots develop flowers. Upon pollination by wind, water, bees, birds, butterflies, lizards, bats, mice, rats and monkeys[74], pollen is transported from the anthers to the stigma. The pollen tube, which emerges from the pollen grain, grows though the style. The sperm that are released from the pollen tube fertilize the egg in the ovule, and the fertilized ovule matures into a seed within the carpel of the flower. The carpel wall expands and either becomes a fleshy fruit[75], which encloses the seeds, and which can be eaten by animals, or dries into an aerodynamic structure. In the former case, the fruit serves as food for animals which then disperse the mature seeds in their feces, and in the latter case, the fruit forms an aerodynamic structure so that the wind can disperse the mature seeds[76,77]. In all cases, the life cycle begins again when the seed germinates[46,78]. Gardeners and botanists have never observed the dispersed seeds migrating back to a flower, the flower shrinking and being reabsorbed into the vegetative tissues until it disappears, the plant shrinking back into the seed, the seed splitting into an egg-containing ovule and a sperm-containing pollen grain, and a pollinator bringing the pollen grain back to an anther. The life cycle always proceeds in one direction in time—never going backwards in time. The asymmetrical order of events in the life cycle of a plant is correlated with the input of radiant energy of the sun, its transformation into electrical, osmotic, and bond energy, and its dissipation as heat[79]. Heat, emitted by the growing plant as long wavelength radiation, flows omnidirectionally away from the plant. A plant never collects infrared radiation from the environment and beams it back to the sun in the form of visible light as it shrinks back into a seed.

When a human being eats a seed, the seed's macromolecular components break down into sugars, fatty acids, and amino acids in the mouth, stomach and small intestine. The sugars and amino acids enter the blood stream and the fatty acids enter the lymphatic system. The nutrients travel to the cells of the body, including those that make up the liver, muscles and brain, to provide energy and mechanical structure. The health food and junk food that we eat today make tomorrow's brain cells and thoughts. The utilization of food always proceeds in this direction[80-86]. The thoughts never congeal into a brain; the nutrients never travel back to the small intestine, stomach and mouth; the metabolites never reform into the macromolecules; and the macromolecules never reorganize into a seed that will then pop out of the mouth of a human, be transported from the dining room to the kitchen and back on the plant. These physiological processes always proceed in one direction in time—never reversing their order in time. The asymmetrical order of events in the life of a human is correlated with the input of potential energy in the form of food, its transformation into electrical, osmotic, and bond energy, and its dissipation as heat[87]. The observations of biologists on the fundamental irreversibility of nature is consistent with Newton's[88] conception of an *"absolute, true, and mathematical time,* [which] *of itself, and from its own nature, flows equably without relation to anything external."*

Many physical processes that are fundamentally irreversible, including the swinging of a pendulum through a viscous medium and the fluorescence of natural plant products, including chlorophyll and quinine, were investigated by George Stokes[89-91] and Lord Rayleigh[92]. These physical processes find their analogs in the living cell. Irreversibility in living cells results from the nonvanishing frictional loss of usable energy that occurs in every reaction that takes place, including the movement of an ion or a water molecule through a channel, the change in the conformation of an enzyme that takes place during metabolism, the movement of vesicles through a thixotropic cytoplasm, the sliding of myosin along actin microfilaments or dynein or kinesin along microtubules, the movement of DNA polymerase along the DNA during replication, the movement of RNA polymerase along the DNA during transcription, the movement of ribosomes along mRNA during translation, and the motion of chromosomes moving to the poles during mitosis and meiosis[10,93-108]. In every living process that requires an input of energy, a dissipation of usable energy occurs as a result of frictional interactions between the object in question and its environment, and consequently, usable energy is degraded in part to nonusable thermal energy. Usable energy is only conserved in transformations between potential energy and kinetic energy under idealized circumstances in which friction does not exist. Under real conditions, described by the Second Law of Thermodynamics, conservation of energy only holds when one takes into consideration the dissipation of potential and/or kinetic energy and their transformation into thermal energy[109,110]. It is important to realize that the principle of the conservation of energy, otherwise known as the First Law of Thermodynamics, does not stipulate a conservation of usable energy. According to Isaac Asimov[111], *"it was only with the proper understanding of heat that physics could be made to deal with the actual world (and with life) and not with merely a fictitious world in the mind of the physicist."* The transactional cost of each transformation, in terms of an increase of entropy, is as real as life itself[1].

Biological organisms and their life cycles are considered to be complicated[112] and thus scientists have turned their attention to simple nonliving things, such as elementary particles, fields, strings, and branes, to try to understand the fundamental nature of the physical world. According to Steven Sherwood, *"Physicists, however, prefer the…approach of avoiding overly complex problems and seeking to strip the more tractable ones to their barest essence*[113]." As successful as this technique has been in elucidating the laws of nature, it runs the risk of marginalizing or eliminating fundamental factors in the search for simplicity, thereby making the description of reality simplistic as opposed to simple. In contrast to the conclusions drawn from observations and experiments of fundamental botanical processes, where precursors always precede products and causes always precede effects, modern physics tells us that the fundamental laws of motion are time-reversal-invariant (TRI), and that in reality, natural processes could occur in either direction of time[4-8].

The concept of time-reversal-invariance grew when Ludwig Boltzmann attempted to reduce thermodynamics to the time-reversible laws of Newtonian mechanics, in which molecules were treated as colliding elastic point particles that behaved like billiard balls. Boltzmann's statistical approach to Newtonian mechanics postulated that a transition from any one configuration or microstate to any other configuration or microstate is equally probable[114-122]. The fact that a mechanical system is far more likely to evolve in time from a lower entropy state, characterized by one or a few microstates to a higher entropy state, characterized by many microstates, is solely a result of chance and statistics, and statistically speaking, there is nothing to prevent the same mechanical system from evolving backwards, consistent with Poincaré's recurrence theorem[123,124], from a higher entropy state, characteristic of the typical final or future condition to a lower entropy state, characteristic of the initial or past conditions. According to Boltzmann[114], *"the two directions of time are indistinguishable in the universe, just as there is no up or down in space....a living being in some particular time phase of his individual world will give different names to the time direction toward the most probable states and to the opposite direction—the former will be called the past, the beginning, the latter the future, the end."* Thus Boltzmann considered the Second Law of Thermodynamics, which is asymmetrical with respect to time, to be a law of nature that is only statistically true, but not fundamentally true. Consequently, he concluded that the asymmetry observed in natural processes was a result of accidental initial conditions and not time-asymmetrical fundamental laws. Boltzmann[114] later wrote, that in describing all known natural phenomena, equations should be employed in which the positive and negative direction of time are *"on an equal footing."* According to Max Born[125], *"Irreversibility is therefore a consequence of th*e *explicit introduction of ignorance* [chance and probability] *into the fundamental laws....in reality, however, the world is reversible."*

---
[1] Transactional costs are a fundamental reality. Ronald H. Coase won the 1991 Nobel Prize in Economics for pointing out the reality of transaction costs in doing business (http://www.nobelprize.org/nobel_prizes/economics/laureates/1991/presentation-speech.html). He found that firms would serve no purpose and much legislation would be pointless if transaction costs vanished. John Bogle, in his book, *Enough, True Measures of Money, Business, and Life*; (John Wiley & Sons: New York, NY, USA; 2008) described the reality of transactional costs on Wall Street.

Einstein's[126] discovery of the relativity of simultaneity and his development of the Special Theory of Relativity, in which time itself was not absolute and universal as Newton assumed but local to each observer, had a monumental influence on our concept of time and in establishing the illusion of time. Hermann Minkowski[127] interpreted the Special Theory of Relativity geometrically in terms of spacetime, in which time and space were just axes of a four-dimensional manifold, and he concluded that *"henceforth space by itself, and time by itself, are doomed to fade away into mere shadows, and only a kind of union of the two will preserve an independent reality."* Influenced by th*e* Theory of Relativity, Hermann Weyl[128] wrote that, *"the two essences, space and time, entering our intuition have no place in the world constructed by mathematical physics."* Upon the death of his friend Michele Besso, Einstein[see 129] wrote a letter to Besso's family, saying, *"...for us physicists believe the separation between past, present, and future is only an illusion, although a convincing one."*

In discussing the influence of the Special Theory of Relativity on our ideas about time, Kurt Gödel[130] wrote,

> *"One of the most interesting aspects of relativity theory for the philosophical-minded consists in the fact that it gave new and surprising insights into the nature of time, of that mysterious and seemingly self-contradictory being which, on the other hand seems to form the basis of the world's and our own existence. The very starting point of special relativity theory consists in the discovery of a new and* *very astonishing property of time, namely the relativity of simultaneity, which to a large extent implies that of succession. The assertion that the events A and B are simultaneous (and, for a large class of pairs of events, also the assertion that A happened before B) loses its objective meaning, in so far as another observer, with the same claim to correctness, can assert that A and B are not simultaneous (or that B happened before A).*

*Following up the consequences of this strange state of affairs one is led to conclusions about the nature of time which are far reaching indeed. In short, it seems that one obtains an unequivocal proof for the view of the philosophers who, like Parmenides, Kant and the modern idealists, deny the objectivity of change and consider change an illusion or an appearance due to a special mode of perception. The argument runs as follows: Change becomes possible only through the lapse of time. The existence of an objective lapse of time, however, means (or, at least, is equivalent to the fact) that reality consists of an infinity of layers of 'now' which come into existence successively. But if simultaneity is something relative in the sense just explained, reality cannot be split up into such layers in an objectively determined way. Each observer has his own set of 'nows,' and none of these various systems of layers can claim prerogative of representing the objective lapse of time."*

Einstein[131] considered Gödel's essay to be "*an important contribution…especially to the analysis of the concept of time.*" According to Rudolf Carnap[132], the fact that the concept of "*now*" "*cannot be grasped by science seemed to* [Einstein] *a matter of painful but inevitable resignation.*"

Building on his knowledge of statistical mechanics and considering the forward and reverse processes that occur at chemical equilibrium, Gilbert N. Lewis[133] proposed a law of entire equilibrium which further diminished the status of "now." Lewis[134] considered equilibrium processes, which are independent of time, and wrote, "*Corresponding to every individual process there is a reverse process, and in a state of equilibrium the average rate of every process is equal to the average rate of its reverse process…..The law of entire equilibrium might have been called the law of reversibility to the last detail. If we should consider any one of the elementary processes which are occurring in a system at equilibrium, and could, let us say, obtain a moving-picture for such a process, then this film reeled backward would present an equally accurate picture of a reverse process which is also occurring in the system and with equal frequency. Therefore in any system at equilibrium, time must lose the unidirectional character which plays so important a part in the development of the time concept. In a state of* *equilibrium, there is no essential difference between backward and forward direction of time, or, in other words, there is complete symmetry with respect to past and future."*

Realizing that *"All the equations of mechanics are equally valid when t is replaced by – t,"* Lewis[135] generalized his inferences to processes that are not independent of time, and wrote, *"throughout the sciences of physics and chemistry, symmetrical or two-way time everywhere suffices….I shall be much disappointed if it can not also be accepted as the statement of a law of physics, of exceptional scope and power, directly applicable to the solution of many classical and modern problems of physics."* Laws that are invariant with respect to the direction of time exhibit time-reversal-invariance or time (T) symmetry.

Emmy Noether discovered that there is a relationship between the symmetry of space and time and the conservation laws. As long as it does not matter how a system is oriented in space, which direction an object moves in space, and when an object moves, then angular momentum, linear momentum, and energy will be conserved. The three conservation laws, which depend on the invariance of spatial rotation, spatial translation and temporal translation, respectively, are all time-reversal-invariant. Eugene Wigner[136] stated in his Nobel lecture that invariance principles are *"touchstones for the validity of possible laws of nature. A law of nature can be accepted as valid only if the correlations which it postulates are consistent with the accepted invariance principles."* Currently, the accepted symmetry of the Standard Theory of Physics is given by the CPT theorem in which any fundamental law must be invariant after changing the sign of the charge (C), the sense of the rotation or parity (P), and the direction of time (T). The fact that the decay of the neutral Kaon is not CP invariant means that its decay is also not T-symmetric[137]. The lack of time symmetry observed in the decay of the neutral Kaon can be looked at as a red herring in the search for the proper description of time in the fundamental time-reversal-invariant laws of physics, or, according to Tim Maudlin[32], *"There is a somewhat better response available. That would be to admit that the laws of physics are not Time Reversal Invariant."* Many of the above arguments for the symmetry of time are presented in a mock debate on whether time exists[138]. In that debate, Tim Maudlin defended the proposition that time does not exist, saying: *"Newton's equations are time reversible…if physics doesn't need it…we don't need it…clear thought shows that time is mere appearance…Einstein space time are combined….why treat one dimension differently from others….moral progress demands it!!!"*

Whether the fundamental laws of physics are symmetrical with respect to time, with reversibility being fundamental; or asymmetrical with respect to time, with irreversibility being fundamental, has been a long-standing topic of debate[139-148]. On the side of symmetry and reversibility, there are people who see mathematics as the essence of reality and believe in the power of simple, elegant and beautiful mathematical formulas to describe and explain the general and fundamental phenomena of nature. In theory, any specific and accidental property of nature can be described and explained by adding additional terms (or epicycles) to the fundamental formula. On the side of asymmetry and irreversibility, there are people who see experience as the essence of reality, and try to find simple mathematical formulas that can describe a complex and complicated reality that includes the ubiquitous, universal, and omnipresent experience of friction, dissipation of usable energy, and entropy. Craig Callender[149] wrote that *"Many, including myself, prefer symmetry and thus TRI laws. But I know of no general way to elevate this aesthetic rationale into an epistemic one."* I assert that all the rationale that has been given for the fundamental nature of time-reversal-invariance is based on aesthetics, and it is possible to derive an irreducible fundamental law of motion, consistent with universal experience, that is asymmetric with respect to time.

Erwin Schrödinger[150], a son of a botanist[151], asked in his book entitled, **What is Life?**, *"When does a physical system—any kind of association of atoms—display 'dynamical law' (in Planck's meaning)…? Quantum theory has a very short answer to this question, viz. at the absolute zero of temperature….This fact was, by the way, not discovered by theory, but by carefully investigating chemical reactions over a wide range of temperatures and extrapolating the results to zero temperature—which cannot actually be reached. This is Walther Nernst's famous 'Heat-Theorem', which is sometimes, and not unduly, given the proud name of the 'Third Law of Thermodynamics' (the first being the energy principle, the second the entropy principle)."*

As I will show in the next section, since absolute zero is unattainable[152], no process, whether microscopic or macroscopic, will exhibit strictly dynamic behavior that is temperature-independent. Consequently, when the temperature is nonvanishing, the thermal energy term, which is the product of temperature ($T$) and change of entropy ($\Delta S$), will also be nonvanishing. As a result, every physical process, particularly those taking place at velocities approaching the speed of light, will be fundamentally and irreducibly irreversible, and will not be subject to Poincaré-type recurrence. Reversible processes, will only occur in theory, when the velocity vanishes; or the temperature vanishes, which is an impossibility according to the Third Law of Thermodynamics.

**2. Results and Discussion**

The fundamental nature of irreversibility was well known to Walter Gretzky[153], Wayne Gretzky's father, who told the future hockey great, *"Go to where the puck is going, not where it has been."* Can we find an irreducible fundamental law that captures the asymmetrical nature of time and the irreversibility of natural processes? Yes, I have done so by re-evaluating Maxwell's wave equation.

Einstein tried to reformulate Maxwell's equations[154] so that they would take into consideration two inertial frames moving relative to each other at velocity *v*, but his attempts were in vain[155]. This led him to assume that Maxwell's wave equation, as it was written with its single explicit velocity (c), was one of the fundamental laws of physics valid in all inertial frames and that the speed of light was invariant. I have reformulated Maxwell's wave equation so that it takes into consideration the changes in the spatial and temporal characteristics of electromagnetic waves observed when there is relative motion between the inertial frame that includes the source and the inertial frame that includes the observer[156]. My reformulation of Maxwell's wave equation is based on the primacy of the Doppler effect, which is experienced by all waves, as opposed to the primacy of the relativity of space and time. Since, for any solution to the second order wave equation in the form of $\Psi = \Psi_o e^{i(\mathbf{k} \cdot \mathbf{r} - \omega t)}$, the angular wave vector (**k**) and distance (**r**) as well as the angular frequency ($\omega$) and time (*t*) are complementary pairs (**k** · **r**) and ($\omega t$), it is only a matter of convention or taste which members of the pairs (**k**, $\omega$) or (**r**, *t*) one assumes to depend on the relative velocity of the source and observer, and which members of the pairs one assumes to be invariant. Einstein chose **r** and *t* to be velocity-dependent and **k** and $\omega$ to be invariant. Due to the omnipresent experience of the Doppler effect when it comes to water, sound, and light waves, and my experience of the invariance of space and time, I chose **k** and $\omega$ to be velocity-dependent and **r** and *t* to be invariant. The Doppler-based relativistic wave equation, which is based solely on kinematic assumptions[156], is given below in two equivalent forms:

$$\frac{\partial^2 \Psi}{\partial t^2} = cc' \frac{\sqrt{c - v\cos\theta}}{\sqrt{c + v\cos\theta}} \nabla^2 \Psi \qquad (1a)$$

$$\frac{\partial^2 \Psi}{\partial t^2} = cc' \frac{1 - \frac{v}{c}\cos\theta}{\sqrt{1 - \frac{v^2 \cos^2\theta}{c^2}}} \nabla^2 \Psi \qquad (1b)$$

where *v* is the magnitude of the relative velocity of the source and observer; $\theta$ is the angle subtending the velocity vector of the source or the observer and the wave vector originating at the source and pointing toward the observer; c is the speed of light through the vacuum and is equal to $\frac{1}{\sqrt{\varepsilon_o \mu_o}}$, where $\varepsilon_o$ is the electric permittivity and $\mu_o$ is the magnetic permeability of the vacuum; and *c'* is the ratio of the angular frequency ($\omega_{source}$) of the source in its inertial frame to the angular wave number ($k_{observer}$) observed in any inertial frame. When the velocity vector and the angular wave vector are parallel and antiparallel, $\theta = 0$ and $\theta = \pi$ radians, respectively. The following equation is a general plane wave solution to the second order relativistic wave equation given above[156]:

$$\Psi = \Psi_o e^{i(\mathbf{k}_{observer} \cdot \mathbf{r} - \omega_{source} \frac{\sqrt{c - v\cos\theta}}{\sqrt{c + v\cos\theta}} t)} \tag{2}$$

Solving the relativistic wave equation given above for the speed of the wave (c = r/t) results in the following relativistic dispersion relation[156]:

$$c = \frac{\omega_{source}}{k_{observer}} \frac{1 - \frac{v}{c}\cos\theta}{\sqrt{1 - \frac{v^2 \cos^2\theta}{c^2}}} = 2.99 \times 10^8 \text{ m/s} \tag{3}$$

indicating that the speed of light (c) is independent of the velocity of the observer. When *v* vanishes, the source and the observer are in the same inertial frame and $\omega_{source} = k_{source} c$. After replacing $\omega_{source}$ with $k_{source} c$, the above equation transforms into a simple relativistic equation that describes the new relativistic Doppler effect[156]:

$$k_{observer} = k_{source} \frac{1 - \frac{v}{c}\cos\theta}{\sqrt{1 - \frac{v^2 \cos^2\theta}{c^2}}} \tag{4}$$

The above equation that describes the new relativistic Doppler effect differs from Einstein's relativistic Doppler effect equation by having a cosine term in both the numerator and the denominator. The cosine term describes the dependence of the first-order and second-order velocity-dependent spatial and temporal properties of electromagnetic waves on the component of the velocity relative to the angular wave vector. The two cosine terms ensure that the effective velocity between the source and the observer is completely relative and depends only on the source and the observer[156].

The Doppler effect[157] characterizes the changes that occur in the spatial and temporal characteristics of a wave as a function of the relative velocity of the source and the observer. The predicted magnitude of the Doppler effect depends on the relativistic transformation used to describe the relationship between two inertial frames. Christian Doppler utilized the Galilean transformation to describe the velocity-dependent changes in the temporal and spatial characteristics of light waves that occur when the source and observer are in two different inertial frames. Einstein[126] modified the Galilean transformation by including the Lorentz factor. The formula I have proposed for the Doppler effect describes and explains, without invoking the relativity of time, the Ives-Stillwell experiments[158], the relativity of simultaneity[156], why particles with either a charge and/or a magnetic moment are prevented from going faster than the speed of light[159], stellar aberration[160], and the Fizeau experiment[160].

The photon density that exists at a given temperature can be readily determined using Planck's law of blackbody radiation[161] or the Stefan-Boltzmann Law[161]. Given that the Third Law of Thermodynamics states that all temperatures must be greater than absolute zero[152], then any charged particle must move through an environment consisting of photons, whose peak wavelength is given by the Wien distribution law[161]. Any particle with a charge and/or a magnetic moment interacts with the environmental photons. The photons that hit the leading edge ($\theta = \pi$) of the charged particle are blue-shifted as a result of the Doppler effect given above and the photons that hit the trailing edge ($\theta = 0$) of the charged particle are red-shifted as a result of the Doppler effect. Since the linear momenta ($p$) of the photons are inversely related to their wavelengths, the linear momenta of the blue-shifted photons are greater than the linear momenta of the red-shifted photons[159]:

$$p_{blue-shifted} = \frac{h}{\lambda} \frac{1 + \frac{v}{c}}{\sqrt{1 - \frac{v^2}{c^2}}} \tag{5}$$

$$p_{red-shifted} = \frac{h}{\lambda} \frac{1 - \frac{v}{c}}{\sqrt{1 - \frac{v^2}{c^2}}} \tag{6}$$

where h is Planck's constant and $\lambda$ represents the peak wavelength of the background radiation with temperature ($T$).

Consequently, the faster the charged particle goes, the greater will be the push backwards from the blue-shifted photons and the lesser will be the push forward from the red-shifted photons. As a result of the anisotropy of the Doppler effect, the ubiquitous photons will exert a temperature- and velocity-dependent, optomechanical counterforce ($\mathbf{F_{Dopp}}$) on the charged particle. Including this ubiquitous and omnipresent counterforce in Newton's Second Law of Motion[159], we get:

$$\mathbf{F}_{app} + \mathbf{F}_{Dopp} = m\frac{dv}{dt} \qquad (7)$$

where $m$ is the invariant mass of a particle with a charge and/or magnetic moment. The counterforce for an univalently-charged particle is given explicitly by[159]:

$$\mathbf{F}_{Dopp} = -\frac{\sigma_B w^2 e^2 \mu_0}{4\alpha \pi h c^2} T^2 \frac{v^2}{\sqrt{1-\frac{v^2}{c^2}}} \qquad (8)$$

where $\sigma_B$ is the Stefan-Boltzmann constant (5.6704 x $10^{-8}$ J m$^{-2}$ s$^{-1}$ K$^{-4}$), $w$ is the Wien coefficient (2.89784 x $10^{-3}$ m K), $e$ is the elementary charge (1.6022 x $10^{-19}$ C), and $\alpha$ is the fine structure constant (7.2973525698(24) x $10^{-3}$). Equation 8 describes the optomechanical counterforce as being a result of the electromagnetic interaction of the particle with the photons. Equation 8 assumes that the absorbed photons are re-emitted isotropically. However, if the photons are reflected or emitted at the same angle in which they were absorbed, the optomechanical counterforce would be twice as large[159]. As I will show below, the optomechanical counterforce is a frictional or viscous force that results in a dissipation of usable energy and, at constant temperature, the increase of entropy. After combining all the constants, equation 8 becomes[159]:

$$\mathbf{F}_{Dopp} = -1.41 \cdot 10^{-39} \frac{N\,s^2}{m^2 K^2} T^2 \frac{v^2}{\sqrt{1-\frac{v^2}{c^2}}} \qquad (9)$$

Equation 9 shows that the counterforce on a univalently-charged particle is proportional to the square of the absolute temperature. Consequently, the counterforce is ten thousand times greater at 300 K than at 3 K. Thus it should take a greater force to accelerate a particle to velocity $v$ at 300 K than at 3 K, which is a testable prediction[159]. By integrating equation 7 with respect to distance we can transform the vectorial force equation into a scalar energy equation[162-164], and we get:

$$\int_{s_1}^{s_2} \mathbf{F}_{app} ds + \int_{s_1}^{s_2} \mathbf{F}_{Dopp} ds = \int_{s_1}^{s_2} m\frac{dv}{dt} ds \qquad (10)$$

Since $ds = \frac{ds}{dt}dt = vdt$, equation 10 becomes:

$$\int_{s_1}^{s_2} \mathbf{F}_{app} ds + \int_{s_1}^{s_2} \mathbf{F}_{Dopp} ds = \int_{t_1}^{t_2} m\, v(t)\, dv \tag{11}$$

which, after taking the integral on the right hand side, equation 11 becomes:

$$\int_{s_1}^{s_2} \mathbf{F}_{app} ds + \int_{s_1}^{s_2} \mathbf{F}_{Dopp} ds = \frac{1}{2}m(v(t))^2 \Big|_{t_1}^{t_2} \tag{12}$$

The applied force, be it gravitational or electrical, is the negative spatial derivative of the potential energy and the potential energy is the negative integral of the applied force. That is:

$$-\int_{s_1}^{s_2} \mathbf{F}_{app} ds = \Delta PE \tag{13}$$

Consequently, equation 12 becomes:

$$-\Delta PE + \int_{s_1}^{s_2} \mathbf{F}_{Dopp} ds = \frac{1}{2}m(v(t))^2 \Big|_{t_1}^{t_2} \tag{14}$$

Since the right hand side of equation 14 is equal to the change in the kinetic energy, as the particle goes from $s_1(t_1)$ to $s_2(t_2)$, we will write the equation as:

$$-\Delta PE + \int_{s_1}^{s_2} \mathbf{F}_{Dopp} ds = \Delta KE \tag{15}$$

where $\Delta KE$ represents the difference in kinetic energy at two points in time. Since the time-dependence of $\Delta KE$ is quadratic, if $\int_{s_1}^{s_2} \mathbf{F}_{Dopp} ds$ vanishes, the direction of time is not limited to the positive numbers, and

$$-\Delta PE = \Delta KE \tag{16}$$

Since $\Delta PE$ has no time dependence and the time dependence of $\Delta KE$ is quadratic, equation 16 is time-reversible-invariant. This is the fundamental equation of classical mechanics, and it is often given in the form of the Lagrangian ($\mathcal{L}$):

$$\mathcal{L} = KE - PE \tag{17}$$

The Lagrangian form employs D'Alembert's Principle. In such a system, the potential energy represents an actual force and the kinetic energy represents a reversed effective force so that dynamics is reduced to statics and change is measured in a system of generalized coordinates in phase or configuration space[162-164]. The Lagrangian, like Newton's Second Law of Motion, from which it is derived, is time-reversal-invariant[165]. However, when $\int \boldsymbol{F}_{Dopp} ds \neq 0$, which is true for any moving system at any temperature above absolute zero, friction is not a fiction and equations 16 and 17 are too simplistic to be fundamental equations of motion. Thus the most reduced and fundamental equation of motion becomes:

$$-\Delta PE = \Delta KE - \int \boldsymbol{F}_{Dopp} ds \tag{18}$$

and we see that, at any temperature above absolute zero, the transformation between potential energy and kinetic energy is not conservative. Substituting equation 8 into equation 18 we get:

$$-\Delta PE = \Delta KE + \int 1.41 \cdot 10^{-39} \frac{N\,s^2}{m^2 K^2} T^2 \frac{v^2}{\sqrt{1-\frac{v^2}{c^2}}} ds \tag{19}$$

If we assume that the temperature (*T*) stays constant, after moving the constants outside of the integral, equation 19 becomes:

$$-\Delta PE = \Delta KE + 1.41 \cdot 10^{-39} \frac{N\,s^2}{m^2 K^2} T^2 \int \frac{v^2}{\sqrt{1-\frac{v^2}{c^2}}} ds \tag{20}$$

After performing a Taylor expansion and including the terms up to and including the second order with respect to velocity, we get:

$$-\Delta PE = \Delta KE + 1.41 \cdot 10^{-39} \frac{N\,s^2}{m^2 K^2} T^2 \int v^2 [1 + \frac{v^2}{2c^2}] ds \tag{21}$$

After simplifying, we get:

$$-\Delta PE = \Delta KE + 1.41 \cdot 10^{-39} \frac{N\,s^2}{m^2 K^2} T^2 \int [v^2 + \frac{v^4}{2c^2}] ds \tag{22}$$

We can integrate more easily after changing the variables. Since *ds* = *vdt*, equation 22 becomes:

$$-\Delta PE = \Delta KE + 1.41 \cdot 10^{-39} \frac{N\,s^2}{m^2 K^2} T^2 \int [v^3 + \frac{v^5}{2c^2}] dt \tag{23}$$

We can change the variables again. Since $dt = \frac{dv}{a}$, equation 23 becomes

$$-\Delta PE = \Delta KE + 1.41 \cdot 10^{-39} \frac{N\,s^2}{m^2 K^2} T^2 \int \frac{1}{a} [v^3 + \frac{v^5}{2c^2}] dv \quad (24)$$

If we assume that the acceleration $a$ is constant (or slowly varying), we can move it outside the integral, and equation 24 becomes:

$$-\Delta PE = \Delta KE + 1.41 \cdot 10^{-39} \frac{N\,s^2}{m^2 K^2} T^2 \frac{1}{a} \int [v^3 + \frac{v^5}{2c^2}] dv \quad (25)$$

Solving the integral for terms up to and including second order with respect to velocity, we get:

$$-\Delta PE = \Delta KE + 1.41 \cdot 10^{-39} \frac{N\,s^2}{m^2 K^2} T^2 \frac{1}{a} [\frac{(v(t))^4}{4} + \frac{(v(t))^6}{12c^2}] \Big|_{t_1}^{t_2} \quad (26)$$

Assuming that there are no agencies, like those suggested by Macquorn Rankine[166], that can reconcentrate this thermal energy back on the particle, the energy of irreversibility is wholly given by the last term in equation 26. We can equate the energy of irreversibility with the thermal energy that is produced during the transformation between $t_1$ and $t_2$. At constant temperature, we will define the change in thermal energy for an irreversible system as $T\Delta S$, where $\Delta S$ is the change of entropy:

$$T\Delta S = 1.41 \cdot 10^{-39} \frac{N\,s^2}{m^2 K^2} T^2 \frac{1}{a} [\frac{(v(t))^4}{4} + \frac{(v(t))^6}{12c^2}] \Big|_{t_1}^{t_2} \quad (27)$$

Dividing both sides of equation 27 by the temperature ($T$), we get the change of entropy that occurs during a transformation between potential and kinetic energy:

$$\Delta S = 1.41 \cdot 10^{-39} \frac{N\,s^2}{m^2 K^2} T \frac{1}{a} [\frac{(v(t))^4}{4} + \frac{(v(t))^6}{12c^2}] \Big|_{t_1}^{t_2} \quad (28)$$

The fact that all moving systems exist at temperatures greater than absolute zero, as stated by the Third Law of Thermodynamics[152], ensures that no mechanical transformation is conservative, and that the Second Law of Thermodynamics is a fundamental[167-170] and not a statistical[170,171] law of nature. This conclusion contrasts with J. Willard Gibbs'[172] view that, "*The laws of thermodynamics, as empirically determined, express the approximate and probable behavior of systems.*"

Clausius[109] codified in the form of the Second Law of Thermodynamics that, as a result of friction, $\Delta S > 0$ for cyclical processes like the Carnot cycle. According to equation 28, the fact that $\Delta S > 0$ for spontaneous noncyclical processes is explained by the interaction between a moving particle and the environmental photons that results in optomechanical friction. Consequently, $\Delta S$ obtained in equation 28, is in harmony with Clausius' equivalency of entropy and friction, and Zermelo's suggestion of introducing dissipative forces in support of his paradigm of causality over probability[173].

This contrasts with Boltzmann's formulation of entropy in which $\Delta S > 0$ as a result of chance and statistics, and $\Delta S$ is subject to Poincaré's recurrence theorem. While Boltzmann's statistical definition of entropy allows for a recurrence of each instant of time, the interaction-based definition of entropy used here results in the conclusion that every instant of time is unique. While equation 28 applies to a single particle and its environment, the change of entropy that occurs in $10^{23}$ independent particles, the number of particles typically found in systems analyzed with statistical mechanics, can be obtained by multiplying $\Delta S$ obtained in equation 28 by the number of independent particles. We can rewrite equation 18 to show that energy is conserved, consistent with the First Law of Thermodynamics, but usable energy is dissipated, consistent with the Second Law of Thermodynamics. Equation 29 explicitly shows the unification of dynamics and thermodynamics at constant temperature:

$$-\Delta PE = \Delta KE + T\Delta S \qquad (29)$$

When Clausius[174] investigated the amount of heat that a body must receive in order to go from its initial condition to its final condition, he found that the quantity of added heat necessary was not a constant, but depended *"not merely on the momentary state of the body, but also in the way in which it has arrived at that state."* Below I will show that when potential energy is transformed into kinetic energy, the details of the process also depends on the path taken, and specifically, on the temperature of the path taken.

When potential energy (e.g. electrical energy or gravitational energy) is converted into the kinetic energy of a mass at any temperature above absolute zero, $v(t_2) > v(t_1)$ and $a > 0$, consequently, $T\Delta S > 0$ and the potential energy is not conserved in the kinetic energy, but distributed between the kinetic energy and the thermal energy, in a path-dependent manner that depends on the temperature of the radiation through which the particle or particles travel (Figure 1). If we were to reverse the motion of the particle in time, so that the kinetic energy and leftward velocity decreases, $v(-t_2) < v(-t_1)$, $(v(-t_2))^4 < (v(-t_1))^4$, $(v(-t_2))^6 < (v(-t_1))^6$ and $a < 0$, $T\Delta S$ is still greater than zero. Consequently, in the "time reversed" situation, $T\Delta S > 0$ and the kinetic energy, which is less than the original potential energy, is not conserved in the potential energy, but distributed between the potential energy and the thermal energy in a path-dependent manner that depends on the temperature of the radiation through which the particle or particles travel.

**Figure 1.** The movement of an electron in an electric field forward in time (top) from the negatively-charged plate to a positively-charged plate. The potential energy (PE) is transformed into kinetic energy (KE) and thermal energy (TS). The movement of an electron in an electric field backwards in time (bottom) from the positively-charged plate to a negatively-charged plate. The kinetic energy (KE) is transformed into potential energy (PE) and thermal energy (TS). The coupled process is not reversible but irreversible because entropy is produced in both directions. Consequently, the round trip journey does not restore the original potential energy. A portion of the original potential energy is no longer usable and is dissipated as thermal energy and can be quantified as an increase of entropy. Since all masses that are composed of charged particles, potential energy in this figure can be generalized to also represent gravitational energy. Since the optomechanical friction will also result in a dissipation of gravitational energy, we can provide an explanation for Newton's Proposition X, Theorem X in Book Three of the Principia[88], in which he takes resistance into consideration and states: *"the motions of the planets in the heavens may subsist an exceedingly long time."*

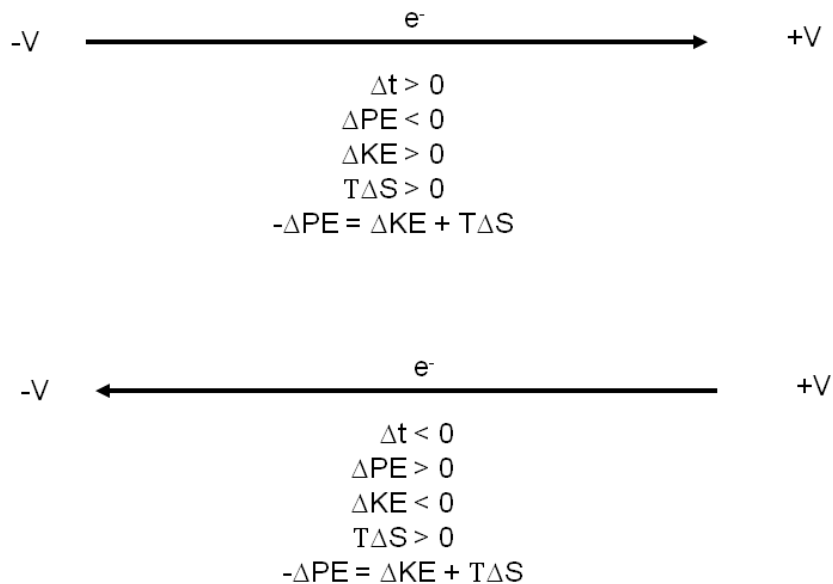

Thus, after running the system forward in time and then backwards in time, the usable energy or the information necessary to rebuild the potential energy is lost and equation 29 predicts an irreversible process that only goes one direction in time. This is the same direction indicated by our intuitive understanding of the arrow of time and the objective arrow of time repeatedly observed in botanical studies.

Equation 29 shows that in any transformation between potential energy and kinetic energy, an irreversible transformation will occur between available energy and unavailable energy, the energy available to perform work and the energy dissipated as thermal energy, or "lost work" at a given temperature and entropy[175]. This is qualitatively consistent with William Thomson's description of natural processes: *"Everything in the material world is progressive. The material world could not come back to any previous state without a violation of the laws which have been manifested to man, that is, without a creative act or an act possessing similar power* [176].*"*

According to Newton's First Law of Motion[88], *"Every body continues in its state of rest, or of uniform motion in a right line, unless it is compelled to change that state by forces impressed on it."* By contrast, equation 30 shows that, in the absence of a potential energy, there will never be uniform motion and the decrease in the kinetic energy of a charged particle over time will depend on the temperature of the surrounding radiation. Uniform motion will only exist and the kinetic energy will only remain unchanged over time, in the absence of a potential energy, consistent with Newton's First Law of Motion, when the temperature is at absolute zero, which is an impossibility according to the Third Law of Thermodynamics[152].

$$\Delta KE = -T\Delta S \qquad (30)$$

At any attainable temperature, $T\Delta S > 0$, consistent with the Second Law of Thermodynamics; consequently, at any attainable temperature, $\Delta KE < 0$ in the absence of an applied force and we see that Newton's First Law of Motion is not a fundamental law of nature. On the other hand, the Second Law of Thermodynamics should be considered to be a primary or fundamental law of nature and not a secondary or statistical law of nature[177].

Consistent with my view of the dynamic processes that occur in plants, the fundamental laws of dynamics should not be time-reversal-invariant. The fact that heretofore the fundamental laws of physics have been deemed time-reversal-invariant, contrary to common sense and intuition came from considering systems to be conservative and sources of resistance, viscosity and friction, including the optomechanical friction that is a consequence of the Doppler effect, to be negligible. The importance of viscous forces was known to Lord Rayleigh[92] when he wrote, "*it is possible that, after all, the investigation in which viscosity is altogether ignored is inapplicable to the limiting case of a viscous fluid when the viscosity is supposed infinitely small.*" The fundamental nature of friction is not only known to biophysical plant cell biologists[10,159], but is also well known to golfers who use a Stimpmeter to measure the green speed[178], and to ballet dancers who rely on the friction between a point shoe and the floor to do a pirouette[179]. In light of my interpretation of mechanics given above, the neutral Kaon decay, whose *"asymmetry in time [is] a result of an interaction, not a boundary condition*[180]*"* gives an insight into the fundamental nature of time, and is not a deviant observation.

In his intuitive and experiential investigation of the nature of the world, Aristotle took friction seriously. Einstein and Infeld[181] wrote in **The Evolution of Physics**, *"The method of reasoning dictated by intuition was wrong and led to false ideas of motion which were held for centuries. Aristotle's great authority throughout Europe was perhaps the chief reason for the long belief in this intuitive idea. We read in the Mechanics....The moving body comes to a standstill when the force which pushes it along can no longer so act to push it....The discovery and use of scientific reasoning by Galileo was one of the most important achievements in the history of human thought, and marks the real beginning of physics. This discovery taught us that intuitive conclusions based on immediate observation are not always to be trusted...."* Einstein and Infeld[181] continued, *"Imagine a road, perfectly smooth, and wheels with no friction at all. Then there would be nothing to stop the cart, so that it would run forever. This conclusion is reached only by thinking of an idealized experiment, which can never be actually performed, since it is impossible to eliminate all external influences. The idealized experiment shows the clew which really formed the foundation of the mechanics of motion."*

In order to create the fundamental physical equations that describe the nature of the physical world, theoretical physicists since Galileo have assumed an idealized world where friction is negligible. In Galileo's[182] **Two New Sciences**, Salivati says to Simplicio on the Fourth Day, *"As to the perturbation arising from the resistance of the medium this is more considerable and does not, on account of its manifold forms, submit to fixed laws and exact description. Thus if we consider only the resistance which the air offers to the motions studied by us, we shall see that it disturbs them all and disturbs them in an infinite variety of ways corresponding to the infinite variety in the form, weight, and velocity of the projectiles. For as to velocity, the greater this is, the greater will be the resistance offered by the air; a resistance which will be greater as the moving bodies become less dense [men gravi]. So that although the falling body ought to be displaced [andare accelerandosi] in proportion to the square of the duration of its motion, yet no matter how heavy the body, if it falls from a very considerable height, the resistance of the air will be such as to prevent any increase in speed and will render the motion uniform; and in proportion as the moving body is less dense [men grave] this uniformity will be so much the more quickly attained and after a shorter fall. Even horizontal motion which, if no impediment were offered, would be uniform and constant is altered by the resistance of the air and finally ceases; and here again the less dense [piu leggier] the body the quicker the process. Of these properties [accidenti] of weight, of velocity, and also of form [figura], infinite in number, it is* *not possible to give any exact description; hence, in order to handle this matter in a scientific way, it is necessary to cut loose from these difficulties; and having discovered and demonstrated the theorems, in the case of no resistance, to use them and apply them with such limitations as experience will teach. And the advantage of this method will not be small; for the material and shape of the projectile may be chosen, as dense and round as possible, so will the spaces and velocities in general be so great but that we shall be easily able to correct them with precision."*

Although Salivati tells Simplicio that it is "*not possible to give any exact description,*" equation 30 shows that, by taking thermodynamics into consideration, we can quantitatively account for the energy lost through optomechanical friction by:

$$T \Delta S = \frac{\sigma_B w^2 e^2 \mu_o}{4 \alpha \pi h c^2} T^2 \frac{1}{a} \left[ \frac{(v(t))^4}{4} + \frac{(v(t))^6}{12c^2} \right] \Big|_{t_1}^{t_2} \tag{31}$$

If equation 31 is correct for a body moving through a distribution of photons characterized by temperature (*T*), Galileo, speaking through Salivati, would only be correct at the absolute zero of temperature, which is unattainable[152]. However, according to Einstein and Infeld[181], "*Galileo's conclusion, the correct one, was formulated a generation later by Newton as the law of inertia. It is usually the first thing about physics which we learn by heart in school....*" It also formed the basis of the Special Theory of Relativity.

At any temperature greater than absolute zero, the idealized world of Galileo, Newton and Einstein dissolves into the real world described by the Laws of Thermodynamics. The concept of optomechanical friction, the dissipation of usable energy, and the increase of entropy must be included in any fundamental law of dynamics. Moreover, the importance of the optomechanical friction becomes greater and greater as the velocity of a particle approaches the speed of light[159], thereby discounting a reason for postulating the symmetry *of* time, and providing evidence in the form of a fundamental law of nature that all things will happen with an asymmetry *in* time.

All particles exist in a radiation field. The partitioning of the potential energy into the kinetic energy of a particle and the environment results in the dissipation of energy due to the interaction of particles with the Doppler-shifted, temperature-dependent blackbody radiation through which they move. This is an irreversible process and the increase in the energy of the environment dissipates away at the speed of light. By decreasing the volume of a space, it could be possible to round up the dissipated usable energy to increase the speed of a particle, but not to restore its original velocity, which is a vector quantity[10].

## 3. Conclusion

The theoretical physicists' conception of the symmetry of time rests on their belief of the fundamental standing of mathematical equations in describing the hidden essence of reality, while the botanists' conception of the asymmetry of time rests on their belief of the fundamental position of experience and experiment in describing the essence of reality. This is similar to Plato's[183] distinction between "being" and 'becoming." To the theoretical physicists, experience and experiment introduce too many trivial, subjective, and accidental properties which obscure the essence of reality, while to the botanists, mathematical equations cannot capture enough of the important parameters, which results in equations that give a simplistic account of the essence of reality. Ernst Mach[184] wrote, *"The object of natural science is the connexion of phenomena; but the theories are like dry leaves which fall away when they have long ceased to be the lungs of the tree of science."* The difference between the botanists' experiential concept of time and the theoretical physicists' mathematical idea of time was captured by Arthur Eddington[185] when he delivered the Messenger Lectures of 1934 at Cornell University.

> *"The view is sometimes held that the dynamic quality of time does not exist in the physical universe and is a wholly subjective impression. Experience presents the physical world as a cinematograph film which is being unrolled in a certain direction; but it is suggested that that is a property of the consciousness, and that there is in the film itself nothing to decide which way it should be unrolled. If this view were right the 'going on of time' ought not to appear in our picture of the external world….In that case, however, we must be careful not to treat the usual past-to-future presentation of the history of the physical universe as truer or more significant that a future-to-past presentation. In particular we must drop the theory of evolution, or at least set alongside it a theory of anti-evolution as equally true and equally significant.*
>
> *If anyone holds this view I cannot answer him by argument; I can only cast aspersions on his character. If he is a professional scientist I say to him: 'You are a teacher and leader whose duty it is to inculcate a true and balanced outlook. But you teach, or without protest allow your colleagues to teach, a one-sided doctrine of evolution. You teach it, not as a colourless schedule of events, but as if there were something significant, perhaps even morally inspiring, in the development out of formless chaos to the richness and adaptation of our present surroundings. Why do you suppress all reference to the sequence from future to past, which according to you is an equally significant sequence to follow? Why do* *you not tell us the story of anti-evolution? Show us how from the diverse species existing to-day Nature anti-evolved clumsier forms, more and more unfitted to survive, till she reached the crudity of Paleozoic life. Show us how from the system of the stars or the planets Nature anti-evolved chaotic nebulae. Narrate the whole story of anti-progress from future to past, and depict the activity of Nature as a force which takes this great work or architecture around us and—makes a hash of it."*

Scientific discoveries have their philosophical consequences[186]. According to Craig Callender[144], *"Ultimately philosophers are interested in non-TRI processes because such processes would appear to pick out a direction of time, in the sense that such a process's evolution in the future would differ from its evolution in the past as a matter of law, and not as a mere matter of fact….The process would seem to be sensitive to time in a fundamental way. As I mentioned, many philosophers would treat this evidence for the anisotropy of time—that there is some basic feature that makes the future and the past differ. A related fact is that many philosophers believe that the non-TRI of the laws may help us explain why the second law of thermodynamics holds."* While the concept of the mathematical symmetry of space and time has been instrumental in the development of theoretical physics and the description of the natural world, perhaps it is limited to the invariance of spatial rotation, space translation, and time translation, three symmetries that have been observed by scientists and lay people in the natural world. On the other hand, the postulate of time-reversal-invariance may be misleading in the search for fundamental laws of physics that apply when $T > 0$, and there may be other, more useful symmetries that can be used to describe the natural world[187]. John Synge[188] *"coined the term R-world to refer to the real world, the world of immense complexity in which we live…."* and the *"M-worlds, with M standing for model or mathematical."* Synge[189] tells us that in science *"there must be two books of rules, one book for the world of reality and the other for the world of mathematics, and that anyone who quotes or uses a rule must know from which of the two books it is taken."* In a lecture given at Clark University in 1899, Ludwig Boltzmann[190] said, *"We shall call an idea about nature false if it misrepresents certain facts or if there are obviously simpler ideas that represent these facts more clearly and especially if the idea contradicts generally confirmed laws of thought…."*

Craig Callender[148] wrote *"The gap between the scientific understanding of time and our everyday understanding of time has troubled thinkers throughout history."* Here I have argued that the scientific understanding of time proffered by the theoretical physicists does not represent the scientific understanding of time subscribed to by all natural scientists. There have been disputes concerning time between physicists and nonphysicists in the past. Frederick Soddy[191] wrote, *"Throughout the latter part of the last century, a controversy as to the possible age of the* *earth as a planet fitted for habitation, existed between two schools, represented by the physicists on the one side and the biologists on the other. Some of the arguments advanced by the former make strange reading at the present time."* John Horgan[192] has recently written in an article in the **Chronicle of Higher Education** about *"how far mainstream physics has drifted from a grounding in empirical science."* Here I have presented my case that the study of natural sciences, including botany, helps one to navigate between the R-world and the M-worlds in the search for the fundamental laws of nature and the nature of time. By analogy with biological processes, I derived the physical basis of irreversibility from readily observable, yet microscopic, quantum mechanical and relativistic physical processes, and related them to the increase of entropy with time, thereby developing a time-irreversible law, in which the time variable *t* cannot sensibly be replaced with *-t*. The definitions of what constitutes a fundamental equation, a law of nature, or a conceptual model have changed over time[190,193,194]. The model that I present here is, in Schrödinger's[2] parlance both a *"visualizable model of the old style"* as well as a *"system of equations and prescriptions as is nowadays favoured."* It represents an intuitive, easily-visualizable, irreducible and fundamental theory, where the algorithm gives different answers when calculating forward in time and backwards in time. It should allow one to not only to analyze the implications of the model but to analyze the presuppositions upon which the model is based.

In the main, the conclusions of modern theoretical physics have supported a philosophy of being as opposed to a philosophy of becoming. Accordingly, Arthur Eddington[195] could write, *"When you say to yourself, 'Every day I grow better and better,' science churlishly replies—'I see no signs of it. I see you extended as a four-dimensional worm in space-time; and although goodness is not strictly within my province, I will grant you that one end of you is better than the other. But whether you grow better or worse depends on which way up I hold you. There is in your consciousness an idea of growth or 'becoming' which, if it is not illusionary, implies that you have a label 'This side up'. I have searched for such a label all through the physical world and can find no trace of it, so I strongly suspect that the label is non-existent in the world of reality."*

The conclusions of botanists are consistent with the idea of irreversibility and the philosophy of becoming[196]. Accordingly, William Thomson[197] wrote, *"The real phenomena of life infinitely transcend human science, and speculation regarding consequences of their imagined reversal is utterly unprofitable."*

I believe that I have made a strong case for the optomechanical cause of irreversibility, the arrow of time, and the philosophy of becoming, and for their place in the primary and fundamental laws of physics. I care about the way we all spend our time, and I hope that you have not wasted the time you spent reading my paper. In celebration of the gift of time and the reality of becoming, I will end this manuscript with quotes from **A Christmas Carol**, written by Charles Dickens[198] and from the poem, **Thanks, Robert Frost**, written by David Ray[199].

> *"Yes! and the bedpost was his own. The bed was his own, the room was his own. Best and happiest of all, the Time before him was his own, to make amends in!*
>
> *'I will live in the Past, the Present, and the Future!' Scrooge repeated, as he scrambled out of bed. 'The Spirits of all Three shall strive within me. Oh, Jacob Marley! Heaven, and the Christmas Time be praised for this! I say it on my knees, old Jacob, on my knees!'"*

Thanks, Robert Frost

> *"Do you have hope for the future?*
> *someone asked Robert Frost, toward the end.*
> *Yes, and even for the past, he replied…."*

**Acknowledgments**

I would like to thank the Cornell University librarians who have preserved the past so that we could influence the future. Andrew Dickson White, the first president of Cornell University, wrote about the Cornell University Library: *"this library will be for generations, nay, for centuries, a source of inspiration to all who would bring good thought of the past to bear in making the future better*[200]."